\documentclass[number,12pt]{elsarticle} 

\usepackage{graphicx}
\usepackage{amssymb}
\usepackage[version=3]{mhchem} 
\usepackage[french]{babel}
\NoAutoSpaceBeforeFDP
\usepackage[latin1]{inputenc}
\usepackage{times}
\usepackage[T1]{fontenc}
\usepackage{textcomp}

\biboptions{square,sort&compress,comma}
\journal{Applied Surface Science}

\begin{document}

\begin{frontmatter}

\title{Optical properties of \ce{Ag}-doped polyvinyl alcohol nanocomposites: 
a statistical analysis of the film thickness effect on the resonance parameters}
\author[]{C. Guyot}
\ead{corentin.guyot@umons.ac.be}
\author[]{M. Vou\'e\corref{cor1}}
\ead{michel.voue@umons.ac.be}
\address{Physique des Matériaux et Optique (LPMO), Université de Mons, \\ 20 
Place du Parc, B-7000 Mons, Belgium}
\cortext[cor1]{Corresponding author}

\begin{abstract}
Nanocomposites made of polymer films embedding silver nanoparticles were prepared by thermal annealing of poly-(vinyl) alcohol films containing \ce{AgNO3}. Low (2.5 \% w:w) and high (25 \% w:w) doping concentration of silver nitrate were considered as well as their effect on the optical properties of thin (30 nm) and thick (300 nm and more) films. The topography and the optical properties (refractive index $n$ and extinction coefficient $k$) of such films were studied by atomic force microscopy and spectroscopic ellipsometry. For a given doping level, the parameters of the surface plasmon-polariton resonance (amplitude, position and width) were shown to be thickness-dependent. Multivariate statistical analysis techniques (principal component analysis and support vector machines) were used to explain the differences in the optical behavior of the thick and thin films.
\end{abstract}


\end{frontmatter}


\section{Introduction}

\label{sec:intro}
Nano-objects and metal nanoparticles (NPs) play a central role in the development of nanotechnology-based optical devices. The optical properties of metal NPs originate from the collective oscillations of their conduction electrons. These are termed localized plasmons or surface plasmon polariton resonances (SPPR) when the free electrons are excited by the electromagnetic field associated to the propagation of light \citep{Kreibig1995,Evanoff2005}. These optical properties are strongly influenced by the shape and size of the NPs but also by the dielectric properties of their environment, in particular when the NPs are embedded in a dielectric matrix such as polymers, silicon dioxide or titanium dioxide layers \citep{Heilmann2005}. A wide range of experimental methods is available for the synthesis of such materials. Besides the methods involving the synthesis of the NPs in a liquid medium, with or without further coating to prevent their aggregation (see e.g. \citep{Kelly2003,Murphy2005,Huang2009,Murphy2011} and the references therein) and their dispersion in a solid phase, NPs can be synthesized \textit{in situ} following the irradiation or the thermal annealing of the solid phase. Although the first approach has an evident advantage of leading to the synthesis of NPs with wide ranges of shapes, size ...,  the synthesis methods of the second category are usually simpler ("one-pot synthesis") and avoid a concentration step of the NPs before their dispersion in the matrix \cite{Vieaud2011}. In particular, in the case of 'soft' matrices such as polymers, the reproducibility of the NPs distributions in size, shape ... remains difficult to achieve and slight variations in the experimental conditions leads to different results evidenced by modifications of the plasmon resonance band parameters. Moreover, the detailed mechanism linking the optical properties of the nanocomposite to their structural parameters is today not fully understood although being the subject of an increasing number of publications. 

The basis principle of these \textit{in situ} synthesis methods is the use of one of the matrix components as a reducing agent of a metal salt. It is well known that polymers are excellent host materials for metal or semiconductor NPs \citep{Mbhele2003} because polymers such as poly-(vinyl) pyrrolidone (PVP) or poly-(vinyl) alcohol (PVA) can not only play the role of reducing agent but also contribute to the stabilization of the NPs. In the latter case, the energy used to promote the reduction of the metal cations originates from a thermal annealing process. In particular, the reduction step of the silver cations uses electrons originating from the \ce{-OH} groups of the PVA molecules

\begin{equation}
\ce{Ag+ + PVA ->[\text{110°C}][\text{60 min}] Ag^0-PVA}
\end{equation}

Interest for such Ag/PVA nanocomposites lays in their optical enhancing properties \citep{Baker20119861,Jiang201494,Whitcomb2008421}, non-linear optical activity \citep{Kuladeep2014,Deng2008911,Anthony2005871,Chen2010223} and antibacterial or anti-fouling properties \citep{Bhat2013,Fuchs2013470,Galya20083178,Thomas20092129}. Some recent articles have been devoted to the study of the role of the polymer matrix \citep{Abargues2009} or of the chemical nature of the metal salt \citep{Yeshchenko2007,Sun2009}.

Usually, the optical properties of plasmonic nanocomposites are probed using reflection or transmission spectrophotometry.  In our previous publications \citep{Dahmouchene2008,Voue2011}, we used spectroscopic ellipsometry (SE) to study the optical properties of Ag-doped PVA (Ag-PVA) films produced by thermal annealing. As explained hereafter, the method has the advantages of being non-destructive and moreover of allowing the simultaneous determination of the film thickness and of its complex refractive index values. To date, only a reduced number of studies have been carried out using SE as analysis technique \citep{Kurbitz2001,Oates2006,Oates2007,Dahmouchene2008,Voue2011}. Excellent review articles have been recently published by Oates, Wormeester and Arwin on spectroscopic ellipsometry studies of plasmon resonances at metal-dielectric interfaces of thin films and metallic nanostructures \citep{Oates2011,Wormeester2013}. In \citep{Dahmouchene2008,Voue2011}, we have shown that, in Ag-PVA films, increasing the annealing time contributed to the onset of more NPs of smaller size and that the intensity of SPPR was correlated to the density of NPs \citep{Voue2011}. In the mid-infrared spectral domain, the presence in the polymer FTIR spectrum of additional peaks located at 1134 cm$^{-1}$ and 1036 cm$^{-1}$ also suggested the presence of bonds between Ag and the polymer matrix \citep{Dahmouchene2008}.

In this article, we report on the influence the film thickness and on the silver doping of the nanocomposite films on the plasmon resonance parameters. More precisely, we compare the resonance parameters in thin and thick films at a given Ag-doping level.

\begin{center}
	\textbf{INSERT HERE FIG. \ref{fig:fig1}}
\end{center}

\section{Materials and methods}

\label{sec:experimental}

\subsection{Nanocomposite film preparation}
\label{sec:nanocomposite}

Polymer films were prepared according to the experimental procedure described in our previous publications \citep{Dahmouchene2008,Voue2011} and based on the original method of Porel and co-workers \cite{Porel2005a,Porel2007}. A 10\% (w:w) poly(vinyl alcohol) solution in MilliQ water was prepared as a stock solution (PVA, MW = 13,000 -- 23,000, 98\% hydrolyzed, Sigma-Aldrich). Dissolution of the polymer was achieved by heating the solution at reflux at 85°C. \ce{AgNO3} (99.99\%, Sigma-Aldrich) was added to the polymer solution after cooling to room temperature to obtain a silver concentration of  2.5\% or 25\%. Percentage corresponds to the w:w silver-to-polymer ratio. Required amount of water was added to the polymer stock solution to obtain 2\% and 8\% polymer solutions. The silver-modified polymer solution was spin-coated on piranha-cleaned (100) silicon wafers fragments (ACM, France). Silicon substrates were chosen because of their high optical contrast with polymer films and their low intrinsic roughness. Depending on the coating conditions, the final thickness of the dried film ranged from 18-25 nm (2\% PVA, 60 sec at 6000 rpm) to 300 nm and above (8\% PVA, 60 sec at 1600 rpm). After being dried in open air atmosphere, the coatings were annealed 60 min at 110°C before their optical properties being measured by spectroscopic ellipsometry (SE) and their topography imaged by atomic force microscopy (AFM). Contrary to the case of glass substrates in which, by increasing the temperature, the coatings became yellowish-brown in color, in the case of silicon substrates, the color of the samples is mainly dominated by the interferometric color and the contribution of the plasmon resonance to the color change can hardly be seen by naked eyes. For comparison, reference PVA films were prepared according to the same method but omitting the \ce{AgNO3} in the polymer solution.

\subsection{Surface characterization}

The topography of the films was studied with a Park XE70 AFM (Park Systems Corp., Korea), operated in air in intermittent contact mode with commercial ACTA tips (resonance frequency: 309 kHz). Areas of 5 $\mu$m x 5 $\mu$m, 2 $\mu $m x 2 $\mu$m and 1 $\mu$m x 1 $\mu$m were imaged for every sample. The resolution of the images was 256 pixels x 256 pixels. The data were processed with the Gwyddion 2.34 SPM images processing software (\emph{http://gwyddion.net/}). The roughness of the samples was characterized by the average surface roughness parameter ($S_a$) and by the root-mean-square surface roughness parameter ($S_q$) defined by 

\begin{equation}
S_a = \frac{1}{MN} \sum_{i=0}^{M-1} \sum_{j=0}^{N-1}{|\,z_{i,j}\,|}
\label{eq:Sa}
\end{equation}

and

\begin{equation}
S_q = \sqrt{\frac{1}{MN} \sum_{i=0}^{M-1} \sum_{j=0}^{N-1}{z_{i,j}^2}}
\label{eq:Rq}
\end{equation}

$M$ and $N$ are length and the width of the image and $z_{i,j}$  is the height of pixel $(i,j)$ above the average plane of the sample \cite{Klapetek2012}.

\subsection{Ellipsometry}

Spectroscopic ellipsometry (SE) is an optical analysis technique based on the change of polarization of light after reflection of an incident light beam on a substrate or on a stratified sample \cite{Azzam1977,Tompkins2005}. Due to its non-destructive nature, SE is a powerful technique to analyze at once the optical properties of materials and the thickness of the different layers of the sample. More specifically, one measures at a given wavelength the ellipticity $\rho$  which is defined by

\begin{equation}
\label{eq:eq1}
\rho = \frac{r_p}{r_s} = \tan \Psi e^{i \Delta}
\end{equation}

where $r_p$ and $r_s$ are the complex reflectance coefficients of the $p-$ ~and $s-$components of the incident light, respectively. $\Psi$ and $\Delta$ are the ellipsometric angles. They are explicitly defined by 

\begin{equation}
\label{eq:eq2}
\tan \Psi =  \frac{|r_p|}{|r_s|} \quad \textrm{and} \quad \Delta = \delta_p - 
\delta_s
\end{equation}

where $\delta_p$ and $\delta_s$ are the phase-shifts undergone by the $p-$ ~ and the $s-$components of the incident wave during the reflection. The angle $\Delta$ therefore measures the relative phase-shift between the perpendicular components of the propagating light wave.

Except in a very limited number of cases, $\Psi$ and $\Delta$ cannot be directly converted into refractive index and/or film thickness. Indeed, the ellipticity is a complex quantity which is function of the angle of incidence, of the optical properties of the materials and of the thickness of the layers. Taking into account the dispersion of the dielectric function, Equation \ref{eq:eq1} has to be inverted in most cases using numerical methods, assuming specular reflection of light at ideal planar interfaces.

The ellipticity of the samples was measured using a SOPRA GES5E rotating polariser spectroscopic ellipsometer with a 1024 channels CCD spectrograph allowing fast and accurate measurements in the range 1.37 to 4.96 eV.  The angle of incidence (AOI) was set to 70 deg. and the ellipsometer operated in parallel beam configuration. SE results were processed using the SOPRA Winelli II software.

A one-layer Cauchy model was chosen to represent the optical properties of the Ag-PVA films in the transparent range and a Lorentzian oscillator was added to account for the localized absorption of the plasmon resonance in visible range \citep{Tompkins2005} :

\begin{equation}
\label{eq:eq3}
n_{PVA}(\lambda) = A_{PVA} + \frac{B_{PVA}}{\lambda^2} \quad \textrm{and} 
\quad k_{PVA}(\lambda) = 0
\end{equation}

The specific contribution of the plasmon resonance to the wavelength-dependent complex dielectric function $\epsilon(\lambda)$ is given by

\begin{equation}
\epsilon(\lambda) = \epsilon_r(\lambda) + i \epsilon_i(\lambda)
\end{equation}

\begin{equation}
\label{eq:eq4}
\epsilon_r(\lambda) = \epsilon_\infty + \frac{A \lambda^2 \left ( \lambda^2 - 
\Lambda_0^2 \right )}{\left ( \lambda^2 - \Lambda_0^2 \right )^2 +\Gamma_0^2 
\lambda^2}
\end{equation}

\begin{equation}
\label{eq:eq5}
\epsilon_i(\lambda) = \frac{A \lambda^3 \Gamma_0}{\left ( \lambda^2 - \Lambda_
0^2 \right )^2 +\Gamma_0^2 \lambda^2}
\end{equation}

where $\lambda$ denotes the wavelength. $\Lambda_0$ is the resonance wavelength of the oscillator, $A$ its oscillator strength and  $\Gamma_0$  
its full-width at half maximum (FWHM) and is hereafter referenced to as the 'width' of the resonance. $\epsilon_\infty$  represents the contributions to $\epsilon_r$ from the resonances at wavelengths much greater than the measurable frequency range.

\subsection{Multivariate statistical analysis}

The multivariate analysis of the data was achieved using the \textit{R} statistical software \citep{RCoreTeam}. The \textit{FactoMineR} package \citep{FactoMineR} and the \textit{e1071} package \citep{e1071} were used to carry out the principal component analysis (PCA) of the resonance peaks parameters and to carry out the classification tasks, respectively. PCA is a multivariate statistical method pioneered by Pearson \citep{Pearson1901} usually used to reduce the dimension of a data matrix keeping the maximum variance in the reduced variable set \citep{Abdi2010}. Support vector machines (SVM) are statistical classification algorithms used to find the best classifier (i.e. separator) between two sets of (eventually overlapping) data points \citep{Vapnik1995}. Levene test and Kruskal-Wallis post hoc comparisons were done using the \textit{agricolae} package \citep{agricolae2014}.

\section{Results and discussion}

\subsection{Topography of the nanocomposite films}

\begin{figure*}[H]
\centering
\begin{tabular}{ll}
(A) & (B) \\
\includegraphics[totalheight=0.40\textwidth]{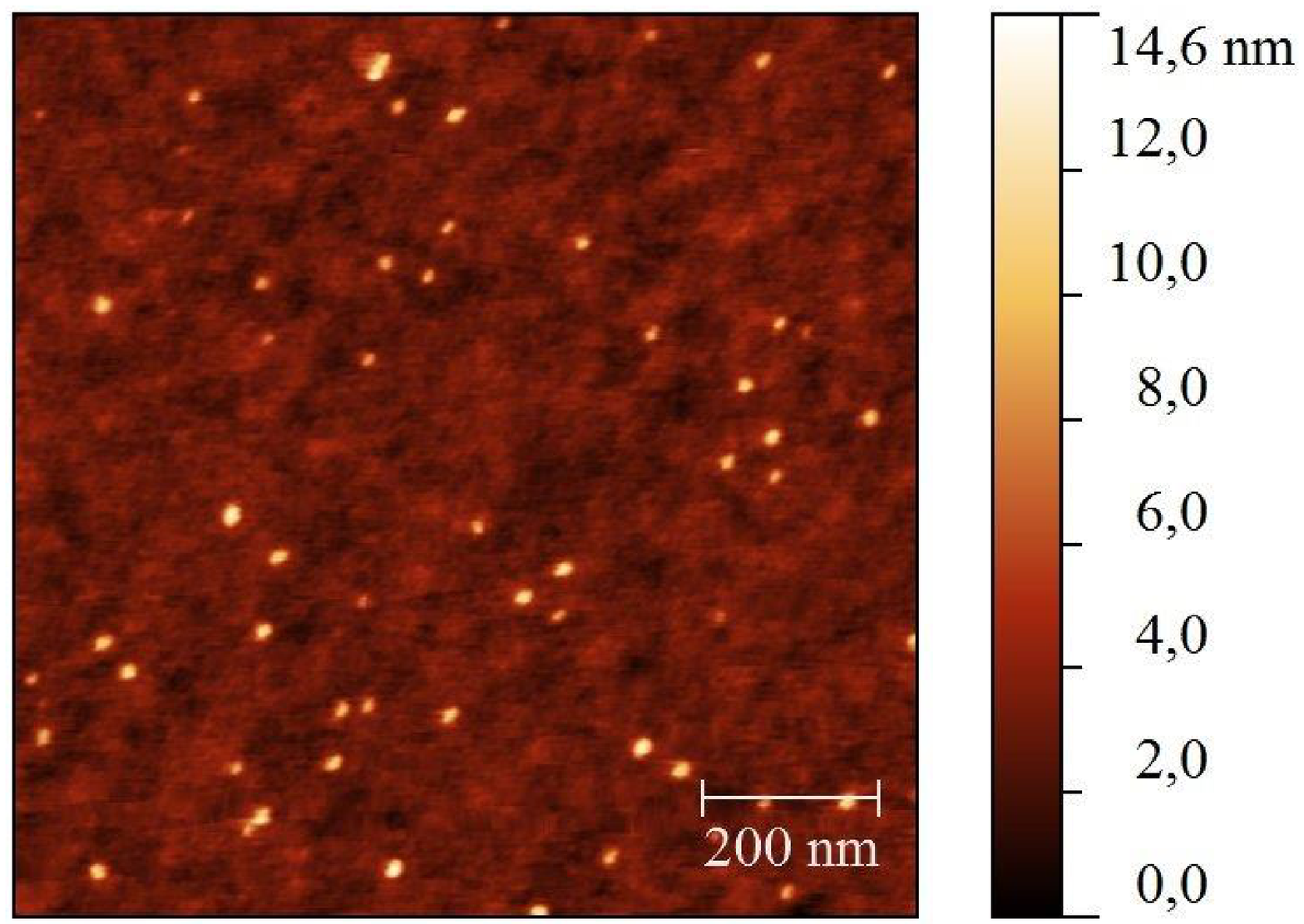} &
\includegraphics[totalheight=0.40\textwidth]{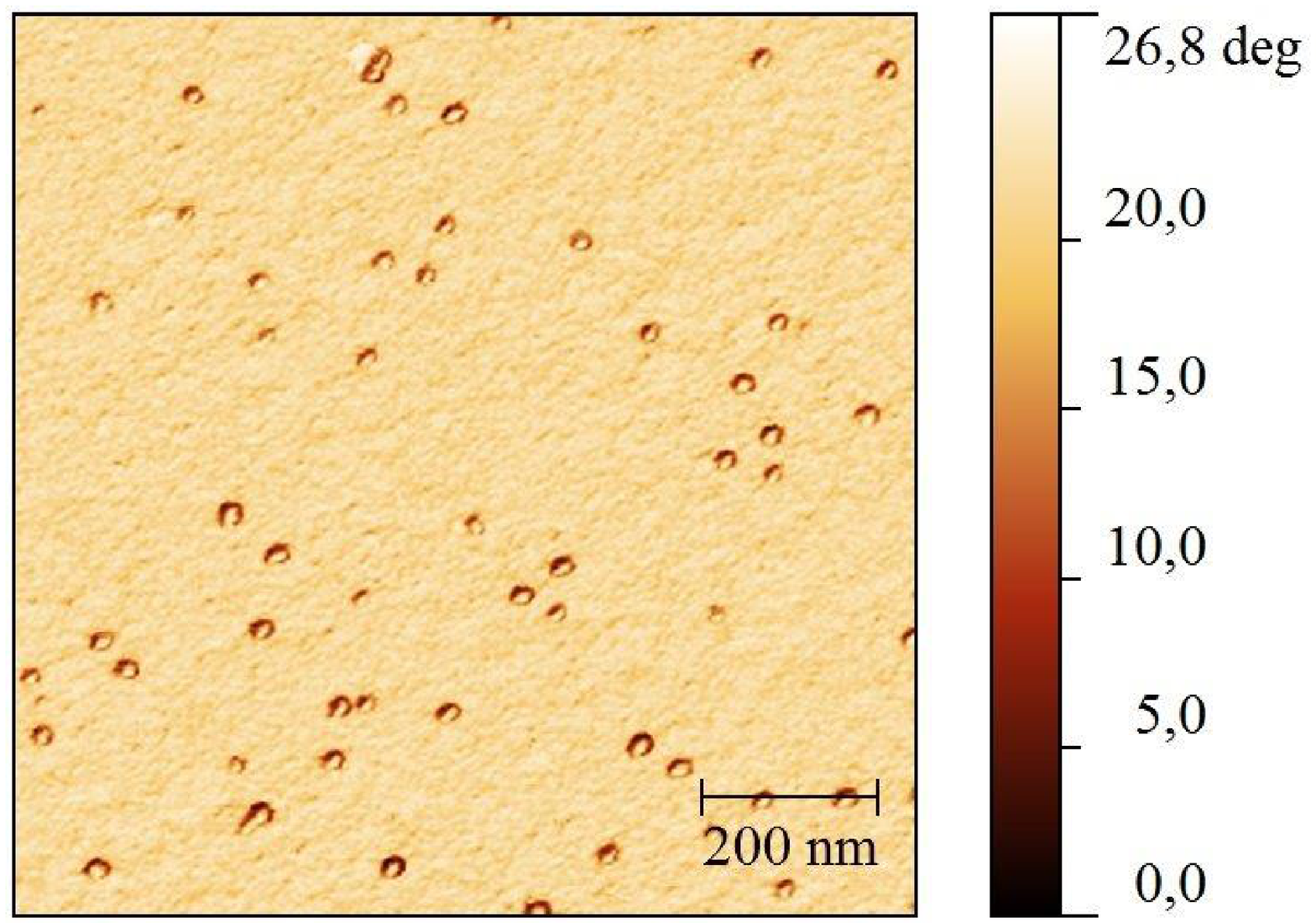} \\
& \\
(C) & (D) \\
\includegraphics[totalheight=0.40\textwidth]{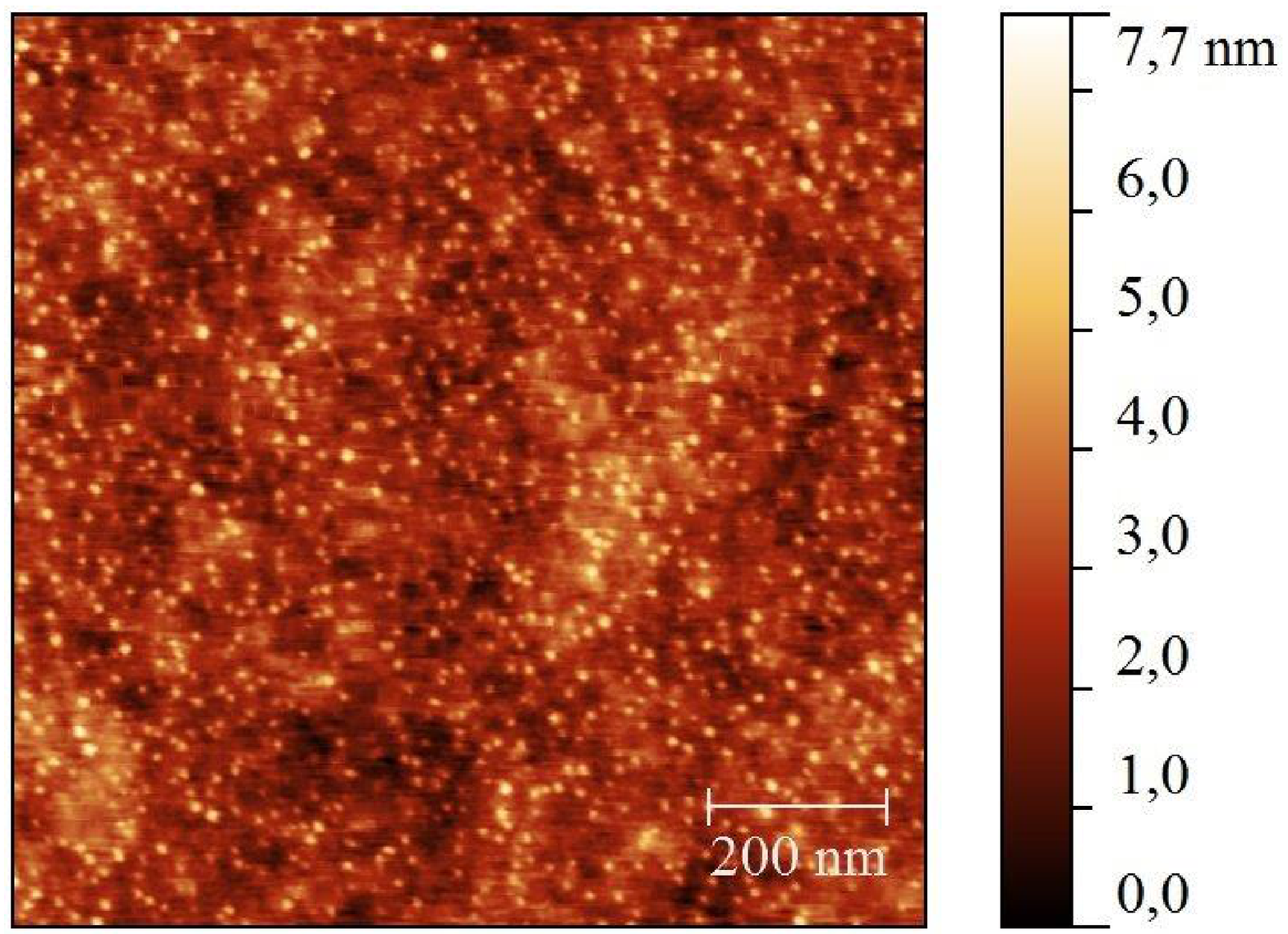} &
\includegraphics[totalheight=0.40\textwidth]{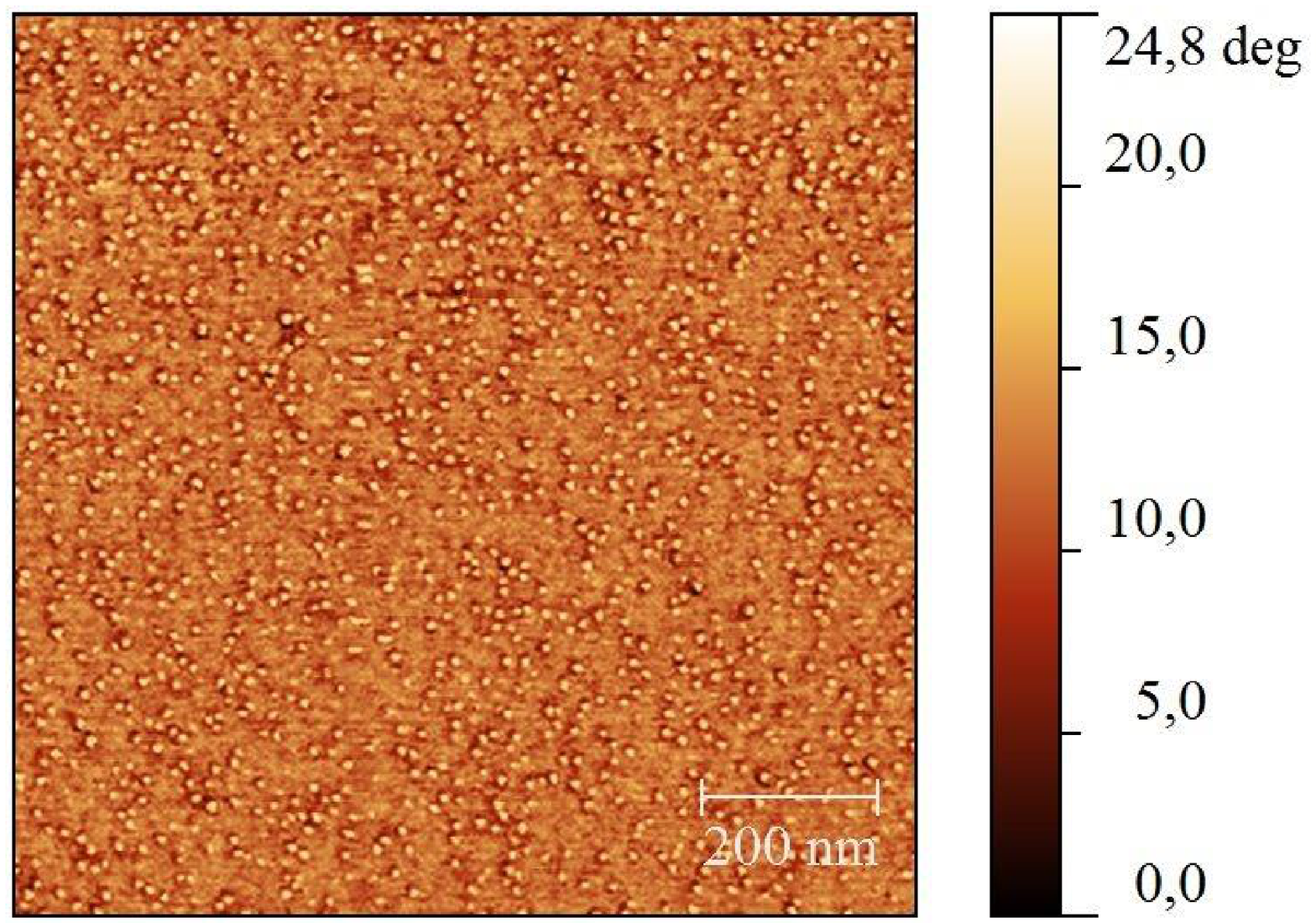} \\
\end{tabular}
\caption{Topography (left) and phase (right) AFM images in intermittent 
contact mode of Ag-PVA film doped with 25\% \ce{AgNO3} (w:w). A,B : $\simeq$ 
30 nm-thick film ; C,D : $\simeq$ 300 nm-thick film. Image size: 1 $\mu$m $
\times$ 1 $\mu$m (256 $\times$ 256 pixels).}
\label{fig:fig1}       
\end{figure*}

To improve our investigation at the nanoscale, the topography of the surface was measured by AFM in intermittent contact mode after the annealing of the film at 110°C during 60 min. Fig. \ref{fig:fig1}A and B represent the topography and phase images (1 $\mu$m $\times$ 1$\mu$m) of a typical 30 nm-thick film. The ratio of \ce{Ag}/PVA was 25\% (w:w). The nanoparticles of silver were clearly observed by a variation of the phase value and a local variation of the height. The first observation of these images shows that the size of the particles is on average $10$ nm. In Fig. \ref{fig:fig1}C and D, we represented the topography and the phase images corresponding to a 300 nm-thick film at the same \ce{Ag}/PVA ratio. The size of nanoparticles seems to be smaller (on average $5-6$ nm). 

\begin{figure}[H]
\centering
\includegraphics[totalheight=0.8\textwidth,angle=-90]{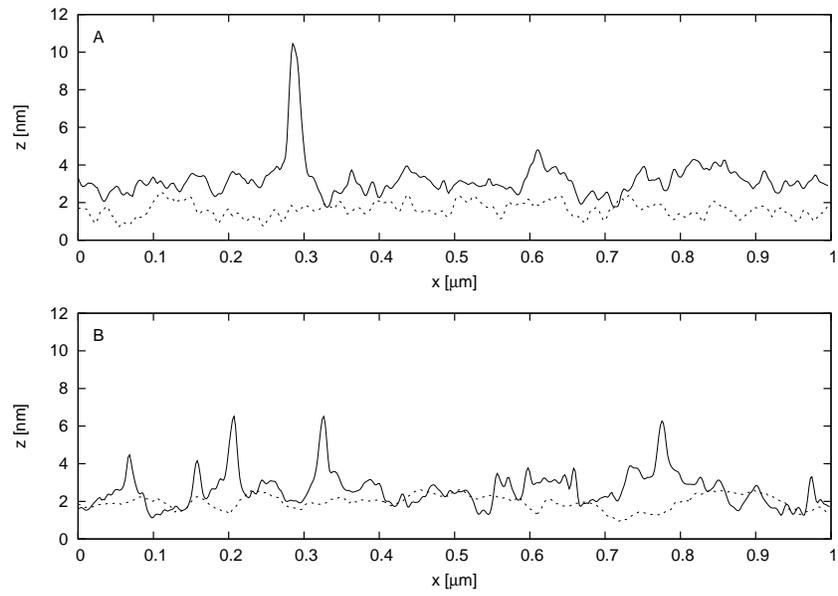}
\caption{Topographic profiles : (A) thin film (B) thick film (Plain line: 
highly doped samples ; Dashed line : undoped PVA sample)}
\label{fig:profils}       
\end{figure}

$S_a$ and $S_q$ were calculated with Gwyddion for doped and non-doped representative samples (Table \ref{table:roughness}). The measurements show an significant difference between doped and non-doped film in both cases (thick or thin films). As seen from both the average and the RMS roughness values, the growth of the nanoparticles induced by the annealing of the film controls the roughness of the surface. If the thick and the thin samples are compared to each other either in phase or in topography, one can see that the particles embedded at the surface of the thick sample are smaller than the particles in the thin one. These features are confirmed by examining the topographic profiles of our films. In Fig. \ref{fig:profils}, the dashed and the plain lines correspond to doped and the undoped samples, respectively. In both case, the roughness of the doped surface is higher.

\begin{center}
	\textbf{INSERT HERE TABLE \ref{table:roughness}}
\end{center}

\begin{center}
	\textbf{INSERT HERE FIG. \ref{fig:profils}}
\end{center}

\subsection{Optical properties}

The purpose of this study was to compare the optical properties of thick and thin films embedding silver NPs grown \textit{in situ} from the same polymer solution and to determine whether the organization of the NPs was equivalent in both cases. To that purpose, the complex refractive index was calculated from SE measurements, i.e. from the $\Psi$ and $\Delta$ spectra.

\begin{figure}[H]
\includegraphics[totalheight=\columnwidth,angle=-90]{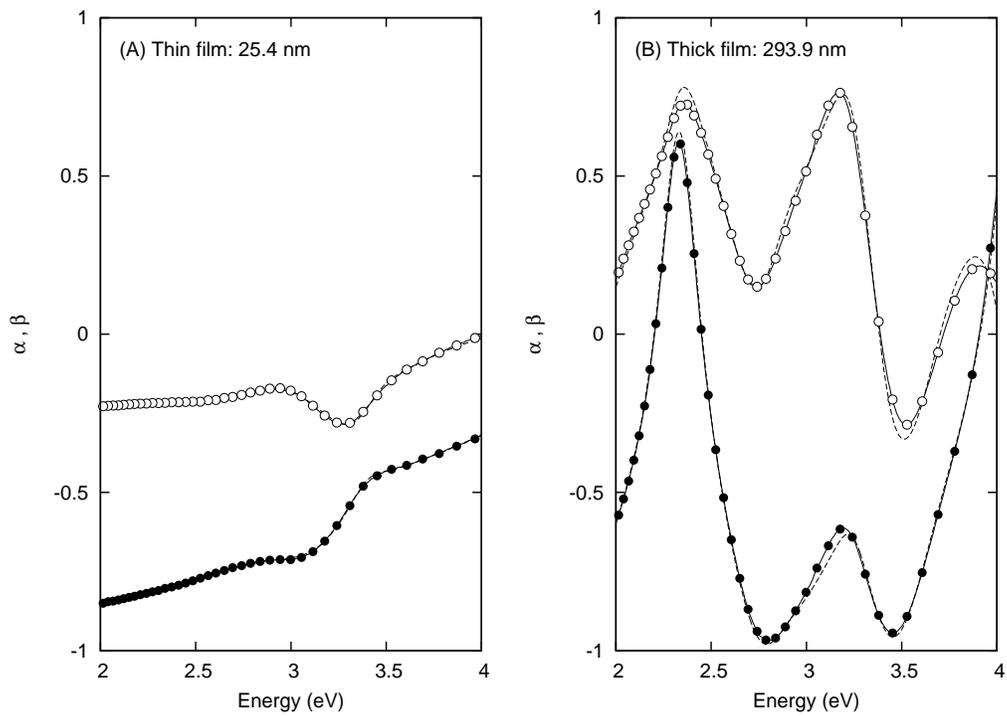}%
\caption[]{%
   Ellipsometric spectra of Ag nanoparticles-doped PVA films (Ag/PVA ratio: 25
\% w:w): A, thin films (thickness: 25.4 nm); B, thick film (thickness: 293.9 
nm). Experimental data: $\alpha = \cos (2 \Psi)$ (filled circles) and  $\beta 
= \sin(2\Psi) \cos(\Delta)$ (open circles). Dashed lines: optimized results 
from the optical model.}
\label{fig:fig3}
\end{figure}

\begin{center}
	\textbf{INSERT HERE FIG. \ref{fig:fig3}}
\end{center}

One of main advantages of SE over conventional spectrophotometric methods is that it allows a simultaneous determination of the optical properties and of the thickness of the layers and not only the measurement of absorption or of transmission spectra. Fig. \ref{fig:fig3} represents the SE data corresponding to a 25.4 nm-thick film (Fig. \ref{fig:fig3}A) and to a 293.9 nm-thick film (Fig. \ref{fig:fig3}B). Instead of representing the $\Psi$ and $\Delta$ ellipsometric angles, we have chosen to present the $\alpha = \cos (2\Psi)$ and $\beta = \sin(2\Psi) \cos(\Delta)$ Fourier coefficients, which are closer to the experimental quantities measured by the ellipsometer. With respect to reference films of equivalent thickness, the ellipsometric data present spectra distortion in the 3.0-3.5 eV range. This is attributed to the localized plasmon resonance associated to the silver NPs. This effect is more noticeable for the thick films (Fig. \ref{fig:fig3}B): one can see three peaks in the $\alpha$-spectrum (filled symbols) but the second is considerably damped due to the energy 
absorption at the resonance. The data generated by the optimized optical model are also presented in Fig. \ref{fig:fig3} (dashed lines). The results generated by the optical model are in close agreement with the experimental data: the residual $\chi^2$ values are respectively equal to 2.7 10$^{-4}$ and 2.9 10$^{-3}$ for the thin and thick films.  The higher discrepancy between experiments and modeling for the thick film is due to a slight thickness inhomogeneity of the samples and to a possible distribution of the NPs in the direction perpendicular to the substrate surface that would result is some minor optical anisotropy. This possible cause has not been further considered in this study.

As previously described \citep{Dahmouchene2008}, to model the optical behavior of our samples, a one-layer Cauchy model was chosen to represent the optical properties of the polymer matrix and a Lorentzian oscillator was added to account for the localized absorption of the plasmon resonance in the visible range. The effect of the presence of the Lorentzian oscillator on the model is equivalent to using the Maxwell-Garnett effective medium approximation (MG-EMA) in a view of obtaining the dielectric function of the nanocomposite films \citep{Aspnes1979,Aspnes2013,Wormeester2013}. With this approach, two problems are implicitly solved : the dependence of the dielectric function of the silver NPs and the generalization of the MG-EMA to take into account the size and shape distribution of the NPs. These two problems can easily be solved for spherical NPs \cite{Kreibig1995,Oates2007,Keita2011} but require more complex approaches for non spherical particles (see e.g. \citep{Evanoff2005,Toudert2012,battie2014}). Moreover, these advanced models have to be supported by complementary experiments based e.g. on the quantitative analysis of transmission electron microscopy images to obtain the size distribution of the NPs. Also is avoided the influence of the doping level of the films: at high doping levels, it is expected that the nanocomposite films cannot be considered anymore as diluted solutions of NPs in a polymer matrix. The analysis of the optical response of nanocomposite films containing a high concentration of metal nanoparticles is more relevant of the metal island model \citep{Lekner1989,Bedeaux2002} which has been recently generalized by Wormeester and co-workers \citep{Kooij2002,Wormeester2003}. Adding roughness or void inclusions did not significantly improved the quality of the fits, even for the thin samples. The use of the Cauchy model for the polymer matrix is justified by the monotonous increase of the refractive index of the polymer when going from the infrared to the UV, as expected for dielectric materials far from the fundamental absorption edge.

\begin{figure}[H]
\includegraphics[totalheight=\columnwidth,angle=-90]{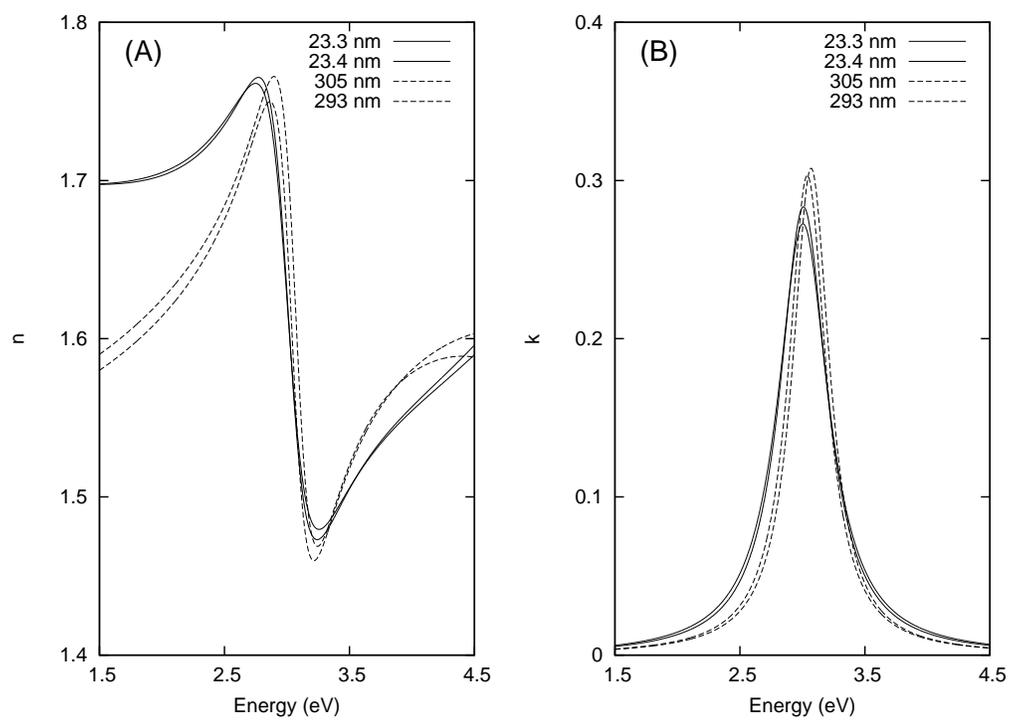}%
\caption{Optical properties of thin (plain lines) and thick (dashed lines) 
silver NPs-doped PVA films (Ag/PVA ratio: 25\% w:w): A, refractive index $n$; 
B, extinction coefficient $k$.}
\label{fig:fig4}
\end{figure}

\begin{center}
	\textbf{INSERT HERE FIG. \ref{fig:fig4}}
\end{center}

The optical properties of the NPs-doped films are represented in Fig. \ref{fig:fig4}A. For each type of film, the refractive index $n$ and the 
extinction coefficient $k$ of two different samples are represented to show the reproducibility of the analysis and indirectly, of the sample 
preparation. The presence of the silver NPs considerably modifies the optical properties in the doped polymer films: the presence of the absorption peak in the 2.5 - 3.5 eV range of the extinction coefficient spectrum (Fig.  \ref{fig:fig4}B) induces a large oscillation in the refractive index value (Fig. \ref{fig:fig4}A) because of the Kramers-Kr\"onig consistency of the optical properties and of the Lorentzian model of the peak. The perturbation of the refractive index occurs over the 1.5 - 4.5 eV range although the absorption peak is limited and the refractive index values are higher than the ones of a pure polymer film (see e.g. Fig. 5 in reference \cite{Voue2011}). The optical properties of thin and thick films are obviously different. The optimized parameters of the plasmon resonance peak are given in Table \ref{table:tab2}. In thin films, the amplitude of the resonance is higher, the position of the resonance smaller and the peaks wider than in thick ones.

\begin{center}
	\textbf{INSERT HERE TABLE \ref{table:tab2}}
\end{center}

Data presented in the \textit{Supporting information} section (Fig. \ref{fig:median}) show that after an annealing time of 40 min. at 110°C the optical properties of the film are stable and that all the \ce{Ag^+} ions are reduced. 

\subsection{Multivariate analysis}

Once the optical properties of the nanocomposites are determined and remembering that they are normally decorrelated from the thickness of the film, we considered the dispersion of the data cloud, the optical response of each nanocomposite being characterized by 3 parameters ($A$, $\Lambda_0$ and $\Gamma_0$). 

Four categorical variables are defined as a function of the thickness of the film ('T' : thick ; 't' : thin) and of the doping level in silver ('H' : high (25\%) ; 'l' : low (2.5 \%)). In the following, these categorical variables will be referred to as the class of the sample.

\subsubsection{Distribution of the resonance parameters}

\begin{figure}[H]
\centering
\includegraphics[width=\columnwidth]{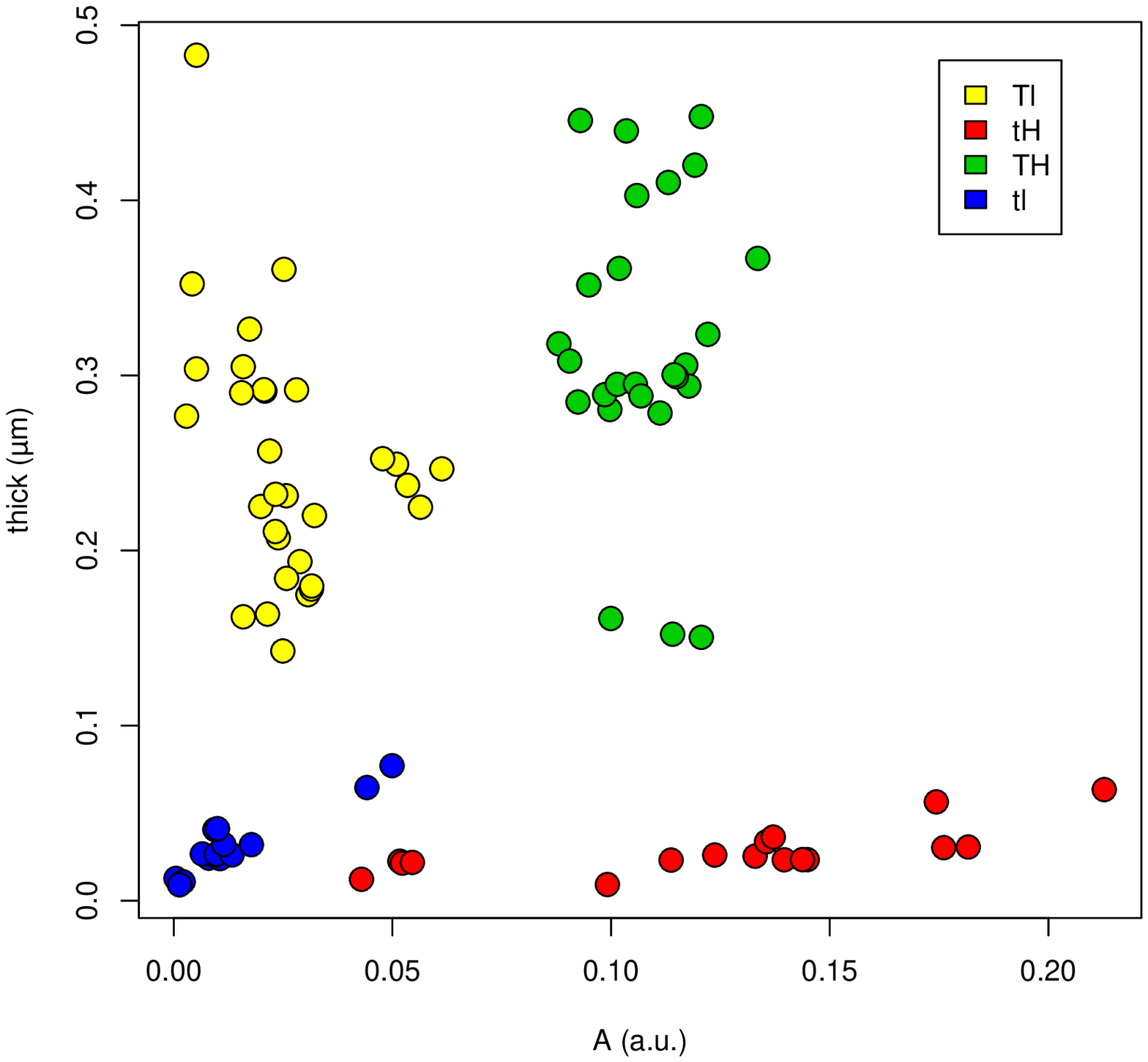}%
\caption{Scatterplot of the thickness of the film (in $\mu$m) versus the strength of the oscillator. Color coding of the categorical variables are given in the legend.}%
\label{fig:scatter}%
\end{figure}

On the basis of the experimental conditions, the 90 samples considered for this study are distributed among 4 classes relatively equivalent sizes. Each of the four categorical variables (i.e. nanocomposite classes) are represented by a different color. Let us first consider the link between the thickness of the film and the strength of the plasmon band $A$ (Fig. \ref{fig:scatter}). One can easily see that the data cloud is subdivided into four distinct classes, each of them being assigned to one type of nanocomposite. 

\begin{center}
	\textbf{INSERT HERE FIG. \ref{fig:scatter}}
\end{center}

\begin{figure}[H]
\centering
\includegraphics[width=\columnwidth]{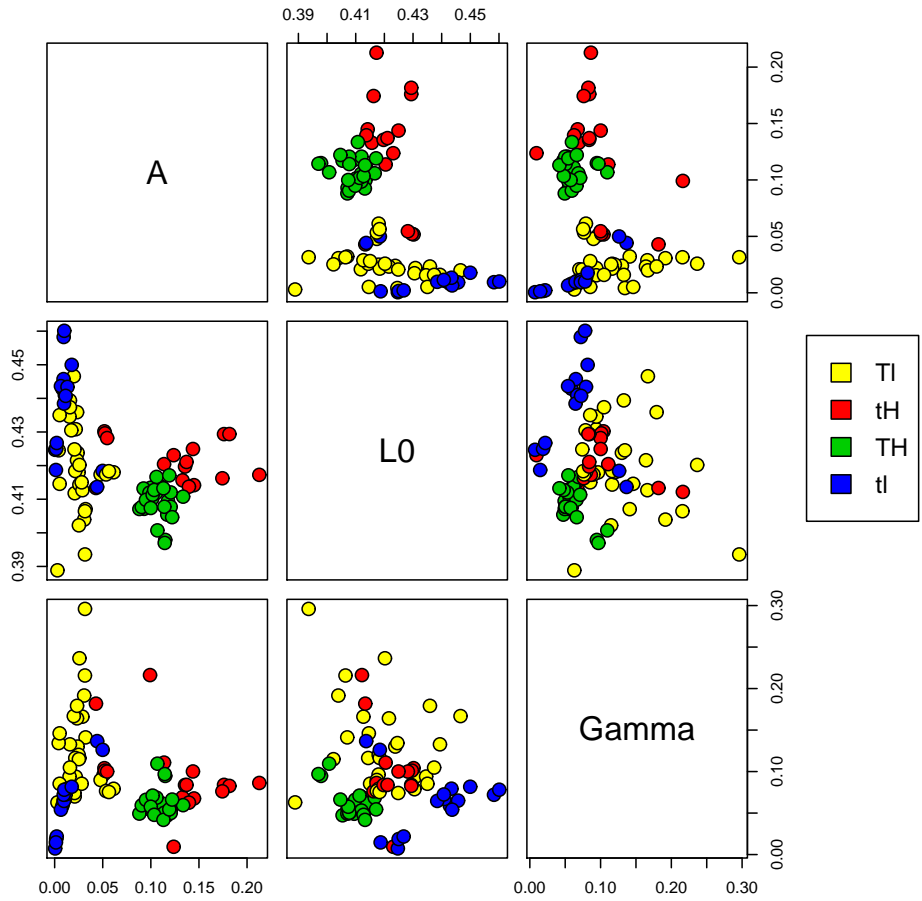}%
\caption{Scatterplot matrix of the resonance parameters. Colour coding of the categorical variables are given in the legend. Units for peak position ("\textit{L0}" or 
$\Lambda_0$) and peak width ("\textit{Gamma}" or $\Gamma_0$) are micrometers.}%
\label{fig:pairs}%
\end{figure}

\begin{center}
	\textbf{INSERT HERE FIG. \ref{fig:pairs}}
\end{center}

As one can see from the scatter plot matrix of the resonance parameters (Fig. \ref{fig:pairs}), the subclouds are much more overlapping, particularly in the case of the $\Gamma_0$ - $\Lambda_0$ plot. In the first row of the scatterplot matrix, different behaviors can be detected between the nanocomposites prepared from solution with a high \ce{AgNO3} concentration (red and green circles) and those prepared from diluted \ce{AgNO3} solution (blue and yellow circles). 

\begin{center}
	\textbf{INSERT HERE TABLE \ref{table:resonance}}
\end{center}

\begin{figure}[H]
\centering
\includegraphics[width=\columnwidth]{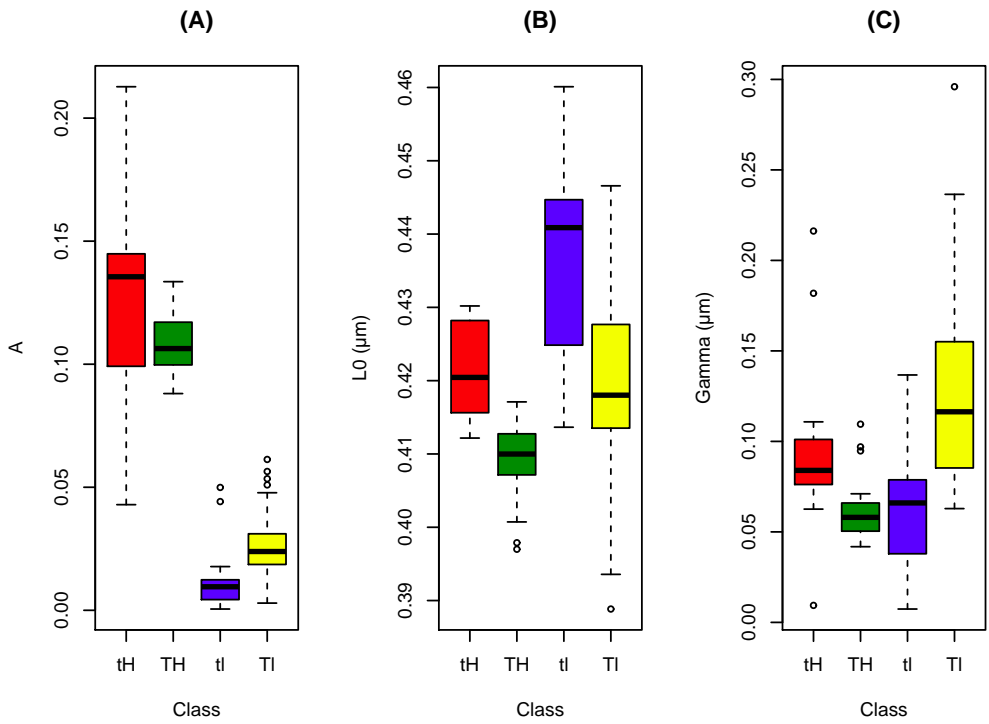}
\caption{Box-and-whiskers plots of the resonance peaks parameters : peak 
height ($A$), peak position ($L0$) and peak width ($Gamma$). Categorical class 
definitions are given in the text.}
\label{fig:boxplots2}       
\end{figure}

This global picture of the data cloud is reinforced by the statistical dispersion of the resonance parameters in terms of box-and-whiskers plots (Fig. \ref{fig:boxplots2}). Heteroscedasticity of the data was checked using a robust version of Levene test. The results show the null hypothesis $H_0$ (equality of all the variances) has to be rejected at a confidence level better than 1\% (data not shown) for all the resonance parameters. Although classical analysis of variance (ANOVA) methods are relatively robust with respect to the non-normal distribution of the variables, differences in variances are more complicated to handle. As a rule of the thumb, classical ANOVA can be used if the ratio of the largest to the smallest variance is < 4. This case is not met by our data. Non parametric tests are therefore required. Results issued from the Kruskal-Wallis test are presented in Table \ref{table:KW}. In a given row of the table, classes represented by the same letter do not statistically differ form each other.
 
\begin{center}
	\textbf{INSERT HERE FIG. \ref{fig:boxplots2}}
\end{center}

As expected, the strength of the oscillator $A$ is strongly dependent on the silver content of the film (Fig. \ref{fig:boxplots2}A). Two trends emerge from the peak position dispersion : increasing the thickness of the films ($t \rightarrow T$) slightly blue-shifts the resonance wavelength, as well as increasing the amount of $\ce{Ag+}$ present in the films ($l \rightarrow H$) (Fig. \ref{fig:boxplots2}B). The evolution of the resonance width is more complicated (Fig. \ref{fig:boxplots2}C). For that reason, the behaviors of the films with low and high silver contents have to be separately described in the remaining parts of this paper. 

\begin{center}
	\textbf{INSERT HERE TABLE \ref{table:KW}}
\end{center}

\subsubsection{Correlation between the plasmon resonance parameters}

\begin{figure}[H]
\centering
\includegraphics[width=0.60\columnwidth]{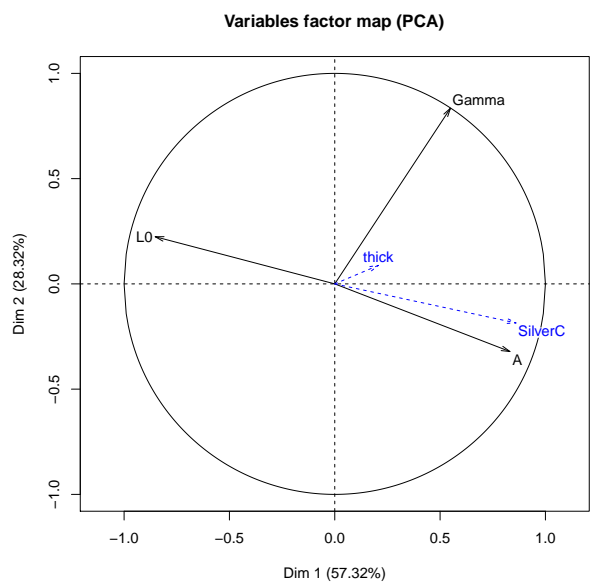}
\includegraphics[width=0.60\columnwidth]{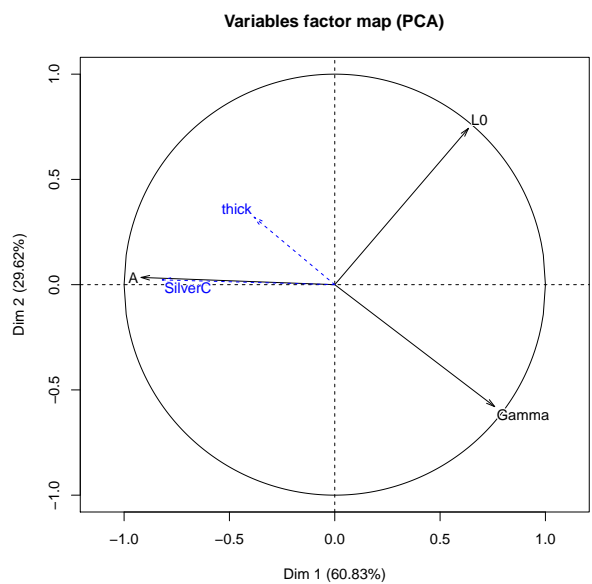}
\caption{Correlation circles in the first factorial plane : (top) thin films 
and (bottom) thick films. Black : Variables used to perfrom the anaysis ; 
Blue : additional projected variables (thickness and siver content of the 
coated solution).}
\label{fig:correlationcircles}       
\end{figure}

To evaluate the correlation between the plasmon resonance parameters ($A$, $L_0$ and $Gamma$) and their link with the experimental variables (thickness of the film and silver content), we performed a principal components analysis to find the principal axes of the data cloud. Thin and thick films were separately considered. Taking into account the reduced number of active variables (3), the percentage of the variance expressed in the first factorial plan is high (Table \ref{table:eigenvalues}). Additional variables (thick : film thickness and SilverC : silver concentration) are also projected in the first factorial plan and are represented in blue in Fig. \ref{fig:correlationcircles}. For both the thin and the thick films, the silver concentration is, as expected, strongly  correlated to the strength of the oscillator $A$. More surprising is the fact that the thickness is not correlated with the plasmon resonance parameters for thin films (Fig. \ref{fig:correlationcircles}, top) and appears clearly to be anti-correlated with the width of the resonance for the thicker ones (Fig. \ref{fig:correlationcircles}, bottom).

\begin{center}
	\textbf{INSERT HERE TABLE \ref{table:eigenvalues}}
\end{center}

\begin{center}
	\textbf{INSERT HERE FIG. \ref{fig:correlationcircles}}
\end{center}

\subsection{Classification in the $\Gamma_0 - \Lambda_0$ plane}

\begin{figure*}[H]
\centering
\includegraphics[totalheight=\textwidth,angle=-90]{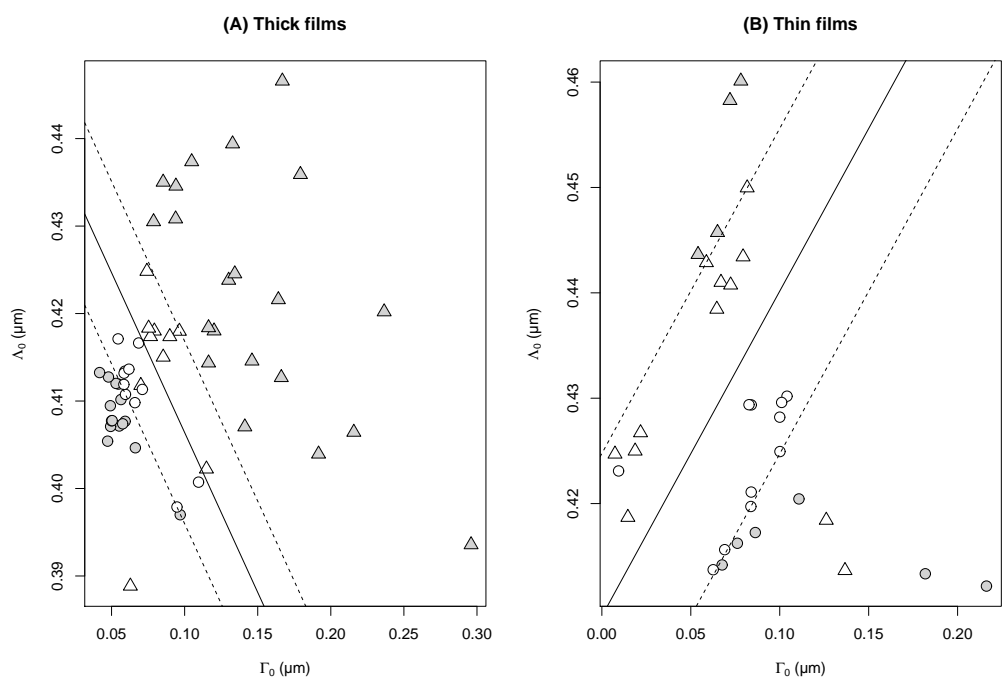}
\caption{Classification of the weakly and highly \ce{Ag}-doped film in the $
\Gamma_0$ -- $\Lambda_0$ plane. (A) Thick films (B) Thin films (Ag-PVA ratio: 
25\% (circles) and 2.5\% (triangles) ; Open symbols: support vectors ; Lines : optimum classifier (plain) and margin (dashed).}
\label{fig:svm1}       
\end{figure*}

As we have seen from the PC analysis, thick and thin films differently behave when we consider the link between the resonance width ($\Gamma_0$) and the position of the plasmonic band ($\Lambda_0$). We therefore attempted to classify the films as a function of the silver content of the coated solution (2.5\% or 25\%). The classification task was performed for the thin and the thick films separately (Fig. \ref{fig:svm1}).

When the data are presented in the $\Gamma_0 - \Lambda_0$ plane, one can see that the set of data points corresponding to the highly doped films (circles) has only a slight overlap with the cloud of the films prepared from the 2.5\% solution (triangles). To perform the classification task in a quantitative way, we used a SVM algorithm \citep{e1071} derived from the original classification algorithm proposed by Vapnik \citep{Vapnik1995}.

SVMs are supervised learning algorithms used to find the best classifier (i.e. separator) between two (or more) data sets, overlapping or not. In the simplest cases (two non-overlapping data clouds in a 2-dimensions space), the algorithm determines the line whose position maximizes the distance to the nearest data points belonging to each class (i.e. the margin). For that reason, SVMs are also named large margin classifiers. In more complicated cases, non-linear classifiers can be used. One important characteristics of the SVMs has to be kept in mind : the equation of  the separator can only be explicitly determined in the case of a linear separator. In other cases, the equation of the separator is not known because the classification task is performed is a derived space (usually of a higher dimension than the data space) on the basis of an inner product only \citep{Vapnik1995}. Besides the choice of the kernel (linear or not), the classifier has to be tuned with respect to the penalty applied when an event is misclassified. In each case, the classifier has been tuned using a 10$\times$ cross-validation process and varying the penalty $C$ in the range 0.01 to 10. Best results were obtained with a cost $C$ for the misclassification equal to 0.15 and 0.40 for the thin and thick films respectively. Optimum results for the classifiers are represented in Fig. \ref{fig:svm1}. The figure shows that, when applied to our data, SVMs allowed us to discriminate between the high- and low-doping level films in a very efficient way : only 3 films over 33  were misclassified in the thin films case and 3 over 54 in the thick films one. The most striking feature of the classification is the behavior of the classifiers : when comparing panels (A) and (B) in Fig. \ref{fig:svm1}, one can easily see that their slope have opposite signs. The classifiers slopes are respectively -0.36 (thick films) and +0.30 (thin films).

\begin{center}
	\textbf{INSERT HERE FIG. \ref{fig:svm1}}
\end{center}

These different optical behaviors are probably strongly related to the growth process of the NPs in films of different thicknesses. In thin films, NPs grow in a 2D-like matrix while in thick films the NPs rapidly form a 3D structure. In forthcoming papers, we will investigate using AFM the real-time growth of the silver NPs in thin and thick films as well as their final spacial distribution based on electron microscopy results.

\section{Conclusions}

In summary, we have studied the optical properties of nanocomposite containing in situ grown nanoparticles. In particular, the study has been focused on the influence of the thickness of the film and of the silver concentration in the polymer matrix on the optical properties of the nanocomposite. We experimentally observed a different behavior of the thick and thin films as a function of the silver concentration. Starting from spectroscopic ellipsometry results, the statistical distribution of the plasmon resonance parameters has been analyzed using principal component analysis and large margin classifiers. Both techniques lead to the conclusion that at a given \ce{Ag^+} doping level, thin and thick films behave differently, probably due to a different spatial distribution of the NPs in the polymer matrix. 

\section*{Acknowledgments}
The authors thank L. Abdessemed for having contributed to the preparation of the nanocomposite films and R. Zoetemelk (ST Instruments, NL) for fruitful discussions and comments on the analysis of the AFM images. This work is financially supported by the F.N.R.S. (FRFC grant nr 1926111). 

\section*{References}
\bibliographystyle{model1a-num-names}
\bibliography{AgPVA_20131104}

\begin{thebibliography}{51}
\expandafter\ifx\csname natexlab\endcsname\relax\def\natexlab#1{#1}\fi
\providecommand{\bibinfo}[2]{#2}
\ifx\xfnm\relax \def\xfnm[#1]{\unskip,\space#1}\fi
\bibitem[{Kreibig and Vollmer(1995)}]{Kreibig1995}
\bibinfo{author}{U.~Kreibig}, \bibinfo{author}{M.~Vollmer},
  \bibinfo{title}{Optical Properties of Metal Clusters},
  \bibinfo{publisher}{Springer}, \bibinfo{year}{1995}.
\bibitem[{Evanoff and Chumanov(2005)}]{Evanoff2005}
\bibinfo{author}{D.~D. Evanoff}, \bibinfo{author}{G.~Chumanov},
  \bibinfo{journal}{ChemPhysChem} \bibinfo{volume}{6} (\bibinfo{year}{2005})
  \bibinfo{pages}{1221--1231}.
\bibitem[{Heilmann(2003)}]{Heilmann2005}
\bibinfo{author}{A.~Heilmann}, \bibinfo{title}{Polymer Films With Embedded
  Metal Nanoparticles}, \bibinfo{publisher}{Springer}, \bibinfo{year}{2003}.
\bibitem[{Kelly et~al.(2003)Kelly, Coronado, Zhao, and Schatz}]{Kelly2003}
\bibinfo{author}{K.~L. Kelly}, \bibinfo{author}{E.~Coronado},
  \bibinfo{author}{L.~L. Zhao}, \bibinfo{author}{G.~C. Schatz},
  \bibinfo{journal}{Journal of Physical Chemistry B} \bibinfo{volume}{107}
  (\bibinfo{year}{2003}) \bibinfo{pages}{668--677}.
\bibitem[{Murphy et~al.(2005)Murphy, Sau, Gole, Orendorff, Gao, Gou, Hunyadi,
  and Li}]{Murphy2005}
\bibinfo{author}{C.~J. Murphy}, \bibinfo{author}{T.~K. Sau},
  \bibinfo{author}{A.~M. Gole}, \bibinfo{author}{C.~J. Orendorff},
  \bibinfo{author}{J.~Gao}, \bibinfo{author}{L.~Gou}, \bibinfo{author}{S.~E.
  Hunyadi}, \bibinfo{author}{T.~Li}, \bibinfo{journal}{The Journal of Physical
  Chemistry. B} \bibinfo{volume}{109} (\bibinfo{year}{2005})
  \bibinfo{pages}{13857--13870}.
\bibitem[{Huang et~al.(2009)Huang, Neretina, and El-Sayed}]{Huang2009}
\bibinfo{author}{X.~Huang}, \bibinfo{author}{S.~Neretina},
  \bibinfo{author}{M.~A. El-Sayed}, \bibinfo{journal}{Advanced Materials}
  \bibinfo{volume}{21} (\bibinfo{year}{2009}) \bibinfo{pages}{4880--4910}.
\bibitem[{Murphy et~al.(2011)Murphy, Thompson, Chernak, Yang, Sivapalan,
  Boulos, Huang, Alkilany, and Sisco}]{Murphy2011}
\bibinfo{author}{C.~J. Murphy}, \bibinfo{author}{L.~B. Thompson},
  \bibinfo{author}{D.~J. Chernak}, \bibinfo{author}{J.~A. Yang},
  \bibinfo{author}{S.~T. Sivapalan}, \bibinfo{author}{S.~P. Boulos},
  \bibinfo{author}{J.~Huang}, \bibinfo{author}{A.~M. Alkilany},
  \bibinfo{author}{P.~N. Sisco}, \bibinfo{journal}{Current Opinion in Colloid
  \& Interface Science} \bibinfo{volume}{16} (\bibinfo{year}{2011})
  \bibinfo{pages}{128--134}.
\bibitem[{Vieaud(2011)}]{Vieaud2011}
\bibinfo{author}{J.~Vieaud}, \bibinfo{title}{Effective optical properties of
  polymer - gold nanoparticle composite film}, Ph.D. thesis, Universit\'e de
  Bordeaux 1, Bordeaux, France, \bibinfo{year}{2011}.
\bibitem[{Mbhele et~al.(2003)Mbhele, Salemane, van Sittert, Nedeljkovi\'c,
  Djokovi\'c, and Luyt}]{Mbhele2003}
\bibinfo{author}{Z.~H. Mbhele}, \bibinfo{author}{M.~G. Salemane},
  \bibinfo{author}{C.~G. C.~E. van Sittert}, \bibinfo{author}{J.~M.
  Nedeljkovi\'c}, \bibinfo{author}{V.~Djokovi\'c}, \bibinfo{author}{A.~S.
  Luyt}, \bibinfo{journal}{Chemistry of Materials} \bibinfo{volume}{15}
  (\bibinfo{year}{2003}) \bibinfo{pages}{5019--5014}.
\bibitem[{Baker et~al.(2011)Baker, Monti, and Nesbitt}]{Baker20119861}
\bibinfo{author}{T.~Baker}, \bibinfo{author}{O.~Monti},
  \bibinfo{author}{D.~Nesbitt}, \bibinfo{journal}{Journal of Physical Chemistry
  C} \bibinfo{volume}{115} (\bibinfo{year}{2011}) \bibinfo{pages}{9861--9870}.
\bibitem[{Jiang et~al.(2014)Jiang, Li, Zhang, Wang, and Zhou}]{Jiang201494}
\bibinfo{author}{T.~Jiang}, \bibinfo{author}{J.~Li},
  \bibinfo{author}{L.~Zhang}, \bibinfo{author}{B.~Wang},
  \bibinfo{author}{J.~Zhou}, \bibinfo{journal}{Journal of Alloys and Compounds}
  \bibinfo{volume}{602} (\bibinfo{year}{2014}) \bibinfo{pages}{94--100}.
\bibitem[{Whitcomb et~al.(2008)Whitcomb, Stwertka, Chen, and
  Cowdery-Corvan}]{Whitcomb2008421}
\bibinfo{author}{D.~Whitcomb}, \bibinfo{author}{B.~Stwertka},
  \bibinfo{author}{S.~Chen}, \bibinfo{author}{P.~Cowdery-Corvan},
  \bibinfo{journal}{Journal of Raman Spectroscopy} \bibinfo{volume}{39}
  (\bibinfo{year}{2008}) \bibinfo{pages}{421--426}.
\bibitem[{Kuladeep et~al.(2014)Kuladeep, Jyothi, Chakradhar, and
  Rao}]{Kuladeep2014}
\bibinfo{author}{R.~Kuladeep}, \bibinfo{author}{L.~Jyothi},
  \bibinfo{author}{S.~Chakradhar}, \bibinfo{author}{D.~Rao},
  \bibinfo{journal}{Optical Engineering} \bibinfo{volume}{53}
  (\bibinfo{year}{2014}) \bibinfo{pages}{071823}.
\bibitem[{Deng et~al.(2008)Deng, Sun, Wang, Zhang, Ming, and
  Zhang}]{Deng2008911}
\bibinfo{author}{Y.~Deng}, \bibinfo{author}{Y.~Sun}, \bibinfo{author}{P.~Wang},
  \bibinfo{author}{D.~Zhang}, \bibinfo{author}{H.~Ming},
  \bibinfo{author}{Q.~Zhang}, \bibinfo{journal}{Physica E: Low-Dimensional
  Systems and Nanostructures} \bibinfo{volume}{40} (\bibinfo{year}{2008})
  \bibinfo{pages}{911--914}.
\bibitem[{Anthony et~al.(2005)Anthony, Porel, Rao, and
  Radhakrishnan}]{Anthony2005871}
\bibinfo{author}{S.~Anthony}, \bibinfo{author}{S.~Porel},
  \bibinfo{author}{D.~Rao}, \bibinfo{author}{T.~Radhakrishnan},
  \bibinfo{journal}{Pramana - Journal of Physics} \bibinfo{volume}{65}
  (\bibinfo{year}{2005}) \bibinfo{pages}{871--879}.
\bibitem[{Chen et~al.(2010)Chen, Tao, Zou, Zhang, and Wang}]{Chen2010223}
\bibinfo{author}{X.~Chen}, \bibinfo{author}{J.~Tao}, \bibinfo{author}{G.~Zou},
  \bibinfo{author}{Q.~Zhang}, \bibinfo{author}{P.~Wang},
  \bibinfo{journal}{Applied Physics A: Materials Science and Processing}
  \bibinfo{volume}{100} (\bibinfo{year}{2010}) \bibinfo{pages}{223--230}.
\bibitem[{Bhat et~al.(2013)Bhat, Karmakar, and Kothari}]{Bhat2013}
\bibinfo{author}{N.~Bhat}, \bibinfo{author}{N.~Karmakar},
  \bibinfo{author}{D.~Kothari}, \bibinfo{journal}{International Journal of
  Nanoscience} \bibinfo{volume}{12} (\bibinfo{year}{2013})
  \bibinfo{pages}{1350029}.
\bibitem[{Fuchs et~al.(2013)Fuchs, Ritz, P\"utz, Mail\"ander, Landfester, and
  Ziener}]{Fuchs2013470}
\bibinfo{author}{A.~Fuchs}, \bibinfo{author}{S.~Ritz},
  \bibinfo{author}{S.~P\"utz}, \bibinfo{author}{V.~Mail\"ander},
  \bibinfo{author}{K.~Landfester}, \bibinfo{author}{U.~Ziener},
  \bibinfo{journal}{Biomaterials Science} \bibinfo{volume}{1}
  (\bibinfo{year}{2013}) \bibinfo{pages}{470--477}.
\bibitem[{Galya et~al.(2008)Galya, Sedlarik, Kuritka, Novotny, Sedlarikova, and
  Saha}]{Galya20083178}
\bibinfo{author}{T.~Galya}, \bibinfo{author}{V.~Sedlarik},
  \bibinfo{author}{I.~Kuritka}, \bibinfo{author}{R.~Novotny},
  \bibinfo{author}{J.~Sedlarikova}, \bibinfo{author}{P.~Saha},
  \bibinfo{journal}{Journal of Applied Polymer Science} \bibinfo{volume}{110}
  (\bibinfo{year}{2008}) \bibinfo{pages}{3178--3185}.
\bibitem[{Thomas et~al.(2009)Thomas, Yallapu, Sreedhar, and
  Bajpai}]{Thomas20092129}
\bibinfo{author}{V.~Thomas}, \bibinfo{author}{M.~Yallapu},
  \bibinfo{author}{B.~Sreedhar}, \bibinfo{author}{S.~Bajpai},
  \bibinfo{journal}{Journal of Biomaterials Science, Polymer Edition}
  \bibinfo{volume}{20} (\bibinfo{year}{2009}) \bibinfo{pages}{2129--2144}.
\bibitem[{Abargues et~al.(2009)Abargues, Abderrafi, Pedrueza, Gradess,
  Marqu\'es-Hueso, Vald\'es, and Mart\'inez-Pastor}]{Abargues2009}
\bibinfo{author}{R.~Abargues}, \bibinfo{author}{K.~Abderrafi},
  \bibinfo{author}{E.~Pedrueza}, \bibinfo{author}{R.~Gradess},
  \bibinfo{author}{J.~Marqu\'es-Hueso}, \bibinfo{author}{J.~L. Vald\'es},
  \bibinfo{author}{J.~Mart\'inez-Pastor}, \bibinfo{journal}{New Journal of
  Chemistry} \bibinfo{volume}{33} (\bibinfo{year}{2009})
  \bibinfo{pages}{1720--1725}.
\bibitem[{Yeshchenko et~al.(2007)Yeshchenko, Dmitruk, Dmytruk, and
  Alexeenko}]{Yeshchenko2007}
\bibinfo{author}{O.~Yeshchenko}, \bibinfo{author}{I.~Dmitruk},
  \bibinfo{author}{A.~Dmytruk}, \bibinfo{author}{A.~Alexeenko},
  \bibinfo{journal}{Materials Science and Engineering B} \bibinfo{volume}{137}
  (\bibinfo{year}{2007}) \bibinfo{pages}{247--254}.
\bibitem[{Sun et~al.(2009)Sun, Qu, Ji, Meng, Wang, Sun, and Qi}]{Sun2009}
\bibinfo{author}{C.~Sun}, \bibinfo{author}{R.~Qu}, \bibinfo{author}{C.~Ji},
  \bibinfo{author}{Y.~Meng}, \bibinfo{author}{C.~Wang},
  \bibinfo{author}{Y.~Sun}, \bibinfo{author}{L.~Qi}, \bibinfo{journal}{Journal
  of Nanoparticles Research} \bibinfo{volume}{11} (\bibinfo{year}{2009})
  \bibinfo{pages}{1005--1010}.
\bibitem[{Dahmouch\`ene et~al.(2008)Dahmouch\`ene, Copp\'ee, Vou\'e, and
  De~Coninck}]{Dahmouchene2008}
\bibinfo{author}{N.~Dahmouch\`ene}, \bibinfo{author}{S.~Copp\'ee},
  \bibinfo{author}{M.~Vou\'e}, \bibinfo{author}{J.~De~Coninck},
  \bibinfo{journal}{physica status solidi (c)} \bibinfo{volume}{5}
  (\bibinfo{year}{2008}) \bibinfo{pages}{1210--1214}.
\bibitem[{Vou\'{e} et~al.(2011)Vou\'{e}, Dahmouch\`{e}ne, and {De
  Coninck}}]{Voue2011}
\bibinfo{author}{M.~Vou\'{e}}, \bibinfo{author}{N.~Dahmouch\`{e}ne},
  \bibinfo{author}{J.~{De Coninck}}, \bibinfo{journal}{Thin Solid Films}
  \bibinfo{volume}{519} (\bibinfo{year}{2011}) \bibinfo{pages}{2963--2967}.
\bibitem[{Kurbitz et~al.(2001)Kurbitz, Porstendorfer, Berg, and
  Berg}]{Kurbitz2001}
\bibinfo{author}{S.~Kurbitz}, \bibinfo{author}{J.~Porstendorfer},
  \bibinfo{author}{K.~J. Berg}, \bibinfo{author}{G.~Berg},
  \bibinfo{journal}{Applied Physics B - Lasers And Optics} \bibinfo{volume}{73}
  (\bibinfo{year}{2001}) \bibinfo{pages}{333--337}.
\bibitem[{Oates(2006)}]{Oates2006}
\bibinfo{author}{T.~Oates}, \bibinfo{journal}{Applied Physics Letters}
  \bibinfo{volume}{88} (\bibinfo{year}{2006}) \bibinfo{pages}{213115}.
\bibitem[{Oates and Christalle(2007)}]{Oates2007}
\bibinfo{author}{T.~Oates}, \bibinfo{author}{E.~Christalle},
  \bibinfo{journal}{Journal of Physical Chemistry C} \bibinfo{volume}{111}
  (\bibinfo{year}{2007}) \bibinfo{pages}{182--187}.
\bibitem[{Oates et~al.(2011)Oates, Wormeester, and Arwin}]{Oates2011}
\bibinfo{author}{T.~Oates}, \bibinfo{author}{H.~Wormeester},
  \bibinfo{author}{H.~Arwin}, \bibinfo{journal}{Progress in Surface Science}
  \bibinfo{volume}{86} (\bibinfo{year}{2011}) \bibinfo{pages}{328 -- 376}.
\bibitem[{Wormeester and Oates(2013)}]{Wormeester2013}
\bibinfo{author}{H.~Wormeester}, \bibinfo{author}{T.~Oates}, in:
  \bibinfo{editor}{M.~Losurdo}, \bibinfo{editor}{K.~Hingerl} (Eds.),
  \bibinfo{booktitle}{Ellipsometry at the Nanoscale},
  \bibinfo{publisher}{Springer-Verlag}, \bibinfo{year}{2013}, pp.
  \bibinfo{pages}{225--256}.
\bibitem[{Porel et~al.(2005)Porel, Singh, Harsha, Rao, and
  Radhakrishnan}]{Porel2005a}
\bibinfo{author}{S.~Porel}, \bibinfo{author}{S.~Singh}, \bibinfo{author}{S.~S.
  Harsha}, \bibinfo{author}{D.~N. Rao}, \bibinfo{author}{T.~P. Radhakrishnan},
  \bibinfo{journal}{Chemistry of Materials} \bibinfo{volume}{17}
  (\bibinfo{year}{2005}) \bibinfo{pages}{9--12}.
\bibitem[{Porel et~al.(2007)Porel, Venkatram, {Narayana Rao}, and
  Radhakrishnan}]{Porel2007}
\bibinfo{author}{S.~Porel}, \bibinfo{author}{N.~Venkatram},
  \bibinfo{author}{D.~{Narayana Rao}}, \bibinfo{author}{T.~P. Radhakrishnan},
  \bibinfo{journal}{Journal of Nanoscience and Nanotechnology}
  \bibinfo{volume}{7} (\bibinfo{year}{2007}) \bibinfo{pages}{1887--1892}.
\bibitem[{Klapetek(2012)}]{Klapetek2012}
\bibinfo{author}{P.~Klapetek}, \bibinfo{title}{Quantitative Data Processing in
  Scanning Probe Microscopy -- SPM Applications for Nanometrology},
  \bibinfo{publisher}{Elsevier}, \bibinfo{year}{2012}.
\bibitem[{Azzam and Bashara(1977)}]{Azzam1977}
\bibinfo{author}{R.~M.~A. Azzam}, \bibinfo{author}{N.~M. Bashara},
  \bibinfo{title}{Ellipsometry and Polarized Light},
  \bibinfo{publisher}{North-Holland}, \bibinfo{year}{1977}.
\bibitem[{Tompkins and Irene(2005)}]{Tompkins2005}
\bibinfo{editor}{H.~Tompkins}, \bibinfo{editor}{E.~Irene} (Eds.),
  \bibinfo{title}{Handbook of Ellipsometry}, \bibinfo{publisher}{Springer},
  \bibinfo{year}{2005}.
\bibitem[{{R Core Team}(2012)}]{RCoreTeam}
\bibinfo{author}{{R Core Team}}, \bibinfo{title}{R: A Language and Environment
  for Statistical Computing}, \bibinfo{organization}{R Foundation for
  Statistical Computing}, \bibinfo{address}{Vienna, Austria},
  \bibinfo{year}{2012}. \bibinfo{note}{{ISBN} 3-900051-07-0}.
\bibitem[{Husson et~al.(2013)Husson, Josse, Le, and Mazet}]{FactoMineR}
\bibinfo{author}{F.~Husson}, \bibinfo{author}{J.~Josse},
  \bibinfo{author}{S.~Le}, \bibinfo{author}{J.~Mazet},
  \bibinfo{title}{FactoMineR: Multivariate Exploratory Data Analysis and Data
  Mining with R}, \bibinfo{year}{2013}. \bibinfo{note}{R package version 1.25}.
\bibitem[{Meyer et~al.(2014)Meyer, Dimitriadou, Hornik, Weingessel, and
  Leisch}]{e1071}
\bibinfo{author}{D.~Meyer}, \bibinfo{author}{E.~Dimitriadou},
  \bibinfo{author}{K.~Hornik}, \bibinfo{author}{A.~Weingessel},
  \bibinfo{author}{F.~Leisch}, \bibinfo{title}{e1071: Misc Functions of the
  Department of Statistics (e1071), TU Wien}, \bibinfo{year}{2014}.
  \bibinfo{note}{R package version 1.6-2}.
\bibitem[{Pearson(1901)}]{Pearson1901}
\bibinfo{author}{K.~Pearson}, \bibinfo{journal}{Philosophical Magazine}
  \bibinfo{volume}{2} (\bibinfo{year}{1901}) \bibinfo{pages}{559--572}.
\bibitem[{Abdi and Williams(2010)}]{Abdi2010}
\bibinfo{author}{H.~Abdi}, \bibinfo{author}{L.~J. Williams},
  \bibinfo{journal}{Wiley Interdisciplinary Reviews: Computational Statistics}
  \bibinfo{volume}{2} (\bibinfo{year}{2010}) \bibinfo{pages}{433--459.}
\bibitem[{Cortes and Vapnik(1995)}]{Vapnik1995}
\bibinfo{author}{C.~Cortes}, \bibinfo{author}{V.~Vapnik},
  \bibinfo{journal}{Machine Learning} \bibinfo{volume}{20}
  (\bibinfo{year}{1995}) \bibinfo{pages}{273--297}.
\bibitem[{de~Mendiburu(2014)}]{agricolae2014}
\bibinfo{author}{F.~de~Mendiburu}, \bibinfo{title}{agricolae: Statistical
  Procedures for Agricultural Research}, \bibinfo{year}{2014}. \bibinfo{note}{R
  package version 1.1-8}.
\bibitem[{Aspnes et~al.(1979)Aspnes, Theeten, and Hottier}]{Aspnes1979}
\bibinfo{author}{D.~E. Aspnes}, \bibinfo{author}{J.~B. Theeten},
  \bibinfo{author}{F.~Hottier}, \bibinfo{journal}{Phys. Rev. B}
  \bibinfo{volume}{20} (\bibinfo{year}{1979}) \bibinfo{pages}{3292--3302}.
\bibitem[{Aspnes(2013)}]{Aspnes2013}
\bibinfo{author}{D.~E. Aspnes}, in: \bibinfo{editor}{M.~Losurdo},
  \bibinfo{editor}{K.~Hingerl} (Eds.), \bibinfo{booktitle}{Ellipsometry at the
  Nanoscale}, \bibinfo{publisher}{Springer-Verlag}, \bibinfo{year}{2013}, pp.
  \bibinfo{pages}{225--256}.
\bibitem[{Keita and En~Naciri(2011)}]{Keita2011}
\bibinfo{author}{A.-S. Keita}, \bibinfo{author}{A.~En~Naciri},
  \bibinfo{journal}{Physical Review B} \bibinfo{volume}{84}
  (\bibinfo{year}{2011}) \bibinfo{pages}{125436}.
\bibitem[{Toudert et~al.(2012)Toudert, Simonot, Camelio, and
  Babonneau}]{Toudert2012}
\bibinfo{author}{J.~Toudert}, \bibinfo{author}{L.~Simonot},
  \bibinfo{author}{S.~Camelio}, \bibinfo{author}{D.~Babonneau},
  \bibinfo{journal}{Physical Review B - Condensed Matter and Materials Physics}
  \bibinfo{volume}{86} (\bibinfo{year}{2012}) \bibinfo{pages}{045415}.
\bibitem[{Battie et~al.(2014)Battie, En~Naciri, Chamorro, and
  Horwat}]{battie2014}
\bibinfo{author}{Y.~Battie}, \bibinfo{author}{A.~En~Naciri},
  \bibinfo{author}{W.~Chamorro}, \bibinfo{author}{D.~Horwat},
  \bibinfo{journal}{Journal of Physical Chemistry C} \bibinfo{volume}{118}
  (\bibinfo{year}{2014}) \bibinfo{pages}{4899--4905}.
\bibitem[{Lekner(1987)}]{Lekner1989}
\bibinfo{author}{J.~Lekner}, \bibinfo{title}{Theory of Reflection of
  Electromagnetic and Particle Waves}, \bibinfo{publisher}{Martinus Nijhoff
  Publishers}, \bibinfo{year}{1987}.
\bibitem[{Bedeaux and Vlieger(2002)}]{Bedeaux2002}
\bibinfo{author}{D.~Bedeaux}, \bibinfo{author}{J.~Vlieger},
  \bibinfo{title}{Optical Properties of Surfaces}, \bibinfo{publisher}{Imperial
  College Press}, \bibinfo{year}{2002}.
\bibitem[{Kooij et~al.(2002)Kooij, Wormeester, Brouwer, van Vroonhoven, van
  Silfhout, and Poelsema}]{Kooij2002}
\bibinfo{author}{E.~S. Kooij}, \bibinfo{author}{H.~Wormeester},
  \bibinfo{author}{E.~A.~M. Brouwer}, \bibinfo{author}{E.~van Vroonhoven},
  \bibinfo{author}{A.~van Silfhout}, \bibinfo{author}{B.~Poelsema},
  \bibinfo{journal}{Langmuir} \bibinfo{volume}{18} (\bibinfo{year}{2002})
  \bibinfo{pages}{4401--4413}.
\bibitem[{Wormeester et~al.(2003)Wormeester, Kooij, and
  Poelsema}]{Wormeester2003}
\bibinfo{author}{H.~Wormeester}, \bibinfo{author}{E.~S. Kooij},
  \bibinfo{author}{B.~Poelsema}, \bibinfo{journal}{Physical Review B}
  \bibinfo{volume}{68} (\bibinfo{year}{2003}) \bibinfo{pages}{085406}.

\end{thebibliography}

\clearpage
\newpage

\clearpage
\newpage

\clearpage
\newpage

\clearpage
\newpage

\clearpage
\newpage

\clearpage
\newpage

\clearpage
\newpage

\clearpage
\newpage

\clearpage
\newpage

\clearpage
\newpage

\begin{table}[H]
\centering
\begin{tabular}{rrrrr}
  \hline
 & \multicolumn{2}{c}{Thick film} & \multicolumn{2}{c}{Thin film} \\
 & Doped & Undoped & Doped & Undoped\\ 
  \hline
$S_a$ (nm) & 0.66 & 0.40 & 0.59 & 0.32 \\
$S_q$ (nm) & 0.85 & 0.52 & 0.77 & 0.42 \\
  \hline
\end{tabular}
\caption{Roughness parameters of the doped (\ce{Ag}/\ce{PVA} = 25\%) and 
undoped (pure PVA) films}
\label{table:roughness}
\end{table}

\begin{table*}[H]
\centering
  \begin{tabular}{rrrrr}
    \hline
    Sample & d (nm) & $A$ & $\Lambda_0$ (nm) & $\Gamma_0$ (nm) \\
    \hline
	Thin & 23.4 $\pm$ 0.2 & 0.145 $\pm$ 0.6 & 414.2 $\pm$ 0.7 & 67.6 $\pm$ 2.9 \\
	& 25.4 $\pm$ 0.3 & 0.133 $\pm$ 0.5 & 415.6 $\pm$ 0.6 & 69.0 $\pm$ 2.6 \\
	Thick & 305.9 $\pm$ 1.7 & 0.117 $\pm$ 0.2 & 405.4 $\pm$ 0.7 & 47.3 $\pm$ 1.6
 \\
	& 293.4 $\pm$ 1.7 & 0.118 $\pm$ 0.2 & 409.5 $\pm$ 0.6& 49.2 $\pm$ 1.5 \\
  \hline
  \end{tabular}
\caption{Typical parameters of the plasmon absorption peak ($A$: amplitude of the absorption peak; $\Lambda_0$: position of the resonance; $\Gamma_0$: width of the resonance) as a function of the film thickness for highly doped PVA films (Ag/PVA ratio: 25\% w:w). Data correspond to the optical properties presented in Fig. \ref{fig:fig4}.}
\label{table:tab2}
\end{table*}

\newpage
\begin{table*}[H]
\centering
\begin{tabular}{lrrrrrrrrr}
  \hline
 & & \multicolumn{4}{c}{Means} & \multicolumn{4}{c}{Standard deviations} \\
class & $N$ & thick & A & L0 & Gamma & thick & A & L0 & Gamma \\ 
  \hline
tH & 17 & 0.028 & 0.125 & 0.421 & 0.095 & 0.014 & 0.050 & 0.006 & 0.046 \\ 
  TH & 27 & 0.311 & 0.102 & 0.410 & 0.064 & 0.082 & 0.018 & 0.005 & 0.018 \\ 
  tl & 16 & 0.030 & 0.013 & 0.437 & 0.064 & 0.019 & 0.014 & 0.014 & 0.036 \\ 
  Tl & 30 & 0.254 & 0.028 & 0.419 & 0.127 & 0.076 & 0.024 & 0.013 & 0.056 \\ 
   \hline
\end{tabular}
\caption{Class sizes ($N$), means and standard deviations of the resonance parameters as a function of the nanocomposite type. Units for thickness ("\textit{thick}"), peak position ("\textit{L0}" or 
$\Lambda_0$) and peak width ("\textit{Gamma}" or $\Gamma_0$) are micrometers.}
\label{table:resonance}
\end{table*}

\begin{table}[H]
	\centering
	\begin{tabular}{ccccc}
		\hline
		Parameter & \multicolumn{4}{c}{Class} \\
		& tH & TH & tl & Tl \\
		\hline
		A & a & a & b & c \\
		$\Lambda_0$ & b & c & a & b \\
		$\Gamma_0$ & b & c & c & a \\
		\hline
		\end{tabular}
	\caption{Summary of the Kruskal-Wallis statistical test for the resonance parameters. In a given row, classes represented by the same letter do not statistically differ form each other. To be compared with the boxplots in Fig. \ref{fig:boxplots2}}
	\label{table:KW}
\end{table}

\newpage
\begin{table*}[H]
\centering
\begin{tabular}{rrrrr}
  \hline
 Samples & Components & Eigenvalue & Variance (\%) & Cum. variance (\%)\\ 
  \hline
Thick & 1 & 1.82 & 60.83 & 60.83 \\ 
  & 2 & 0.89 & 29.62 & 90.44 \\ 
  & 3 & 0.29 & 9.56 & 100.00 \\ 
   \hline
Thin & 1 & 1.72 & 57.32 & 57.32 \\ 
  & 2 & 0.85 & 28.32 & 85.64 \\ 
  & 3 & 0.43 & 14.36 & 100.00 \\ 
   \hline
\end{tabular}
\caption{Eigenvalues, percentage and cumulative percentage of the variance for the thick and thin films}
\label{table:eigenvalues}
\end{table*}

\clearpage
\newpage
\section*{Supporting information}

\setcounter{figure}{0}
\makeatletter 
\renewcommand{\thefigure}{S\@arabic\c@figure}
\makeatother

\begin{figure*}[H]
\centering
\includegraphics[totalheight=0.8\textwidth,angle=-90]{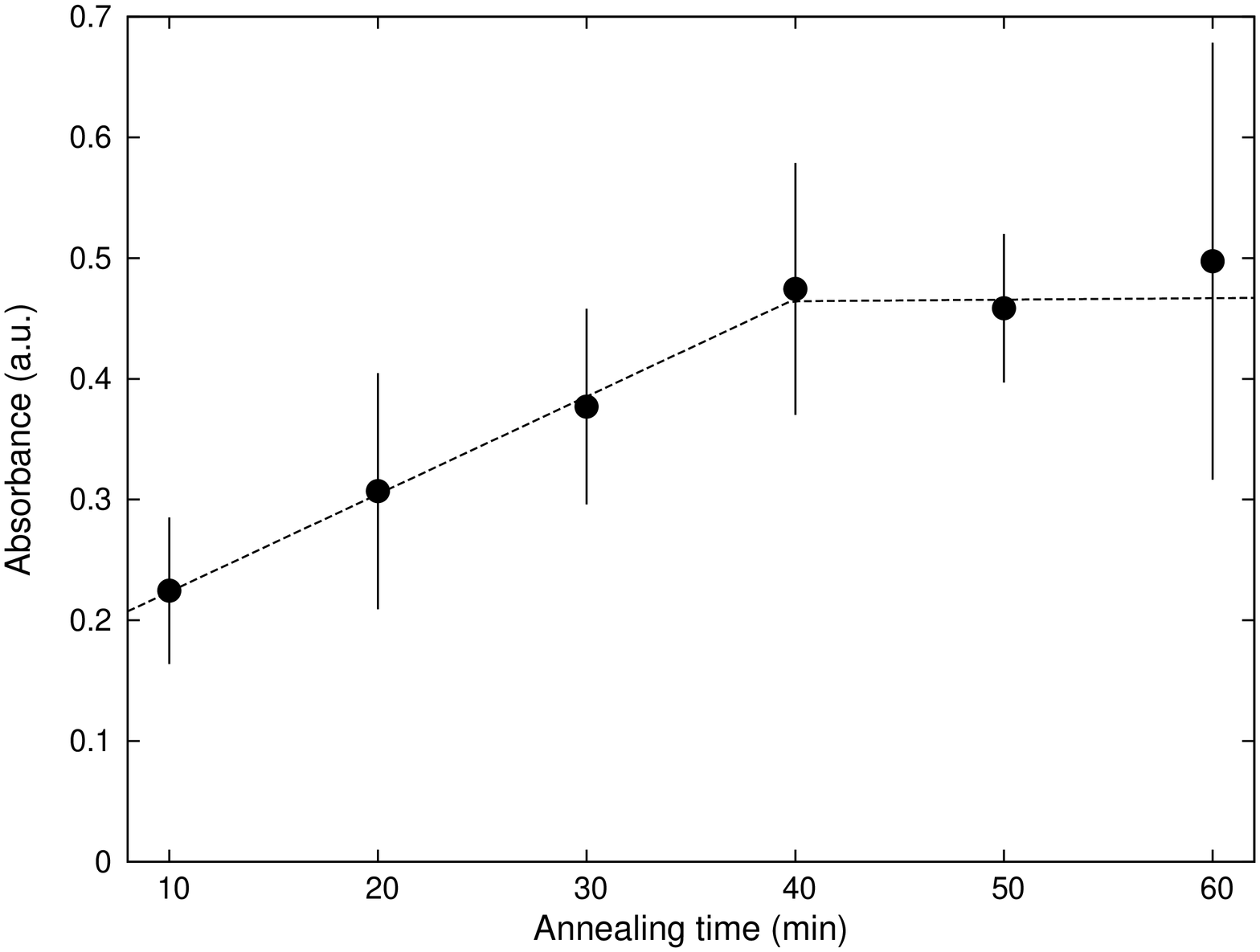}
\caption{Time evolution of the absorbance at peak maximum (Annealing temp. : 110°C -- Data point: median, Error bars: interquartile range, dashed lines: weighed regression lines).}
\label{fig:median}       
\end{figure*}

Figure \ref{fig:median} represents the time evolution of the height of the resonance peak observed for Ag-PVA films on glass substrates during annealing at 110°C. The films were prepared in the same experimental conditions as those used for the 'TH' samples on silicon. The absorbance was measured using a Genesys 20 UV-Vis spectrometer (Thermo, USA). The spectra (data not shown) were recorded from 350 nm to 850 nm. As shown on the figure, the absorbance stabilizes after 40 min., indicating the complete reduction of the \ce{Ag^+} ions. The fluctuations are due to the changes in thickness from one sample to the other. 
 
\end{document}